\documentclass{egpubl}
\usepackage{eurovis2023}
\usepackage{xcolor}
\usepackage{tikz}
\usepackage{multirow}
\usepackage{ulem}
\SpecialIssuePaper         % uncomment for final version of Computer Graphics Forum, special issue
\usepackage[T1]{fontenc}
\usepackage{dfadobe}  

\usepackage{cite}  % comment out for biblatex with backend=biber
\BibtexOrBiblatex
\electronicVersion
\PrintedOrElectronic
\usepackage{egweblnk}
\usepackage{physics}

\newcommand{\toolName}{\textit{VENUS}}

\usepackage[linewidth=0pt,linecolor=blue]{mdframed}
\newcommand{\modify}[1]{\textcolor{black}{#1}}

\usepackage{amsmath}

\usepackage{tikz}
\usepackage{xcolor}
\definecolor{lightGrey}{rgb}{0.94, 0.97, 1.0}

\definecolor{component}{RGB}{89, 89, 89}
\newcommand*\component[1]{\tikz[baseline=(char.base)]{
            \node[shape=circle,draw=component,text=component, thick, inner sep= 1pt] (char) {\textsf{#1}}}}
            
\definecolor{subcomponent}{RGB}{137, 137, 137}
\newcommand*\subcomponent[1]{\tikz[baseline=(char.base)]{
            \node[shape=circle,fill=subcomponent,text=white, inner sep= 0.5pt] (char) {\textsf{\small #1}}}}

\definecolor{subcomponent}{RGB}{137, 137, 137}
\newcommand*\Subcomponent[1]{\tikz[baseline=(char.base)]{
            \node[shape=circle,fill=subcomponent,text=white, inner sep= 1.3pt] (char) {\textsf{\small #1}}}}

\definecolor{redSquare}{RGB}{255, 92, 15}
\newcommand*\redSquare[1]{\tikz[baseline=(char.base)]{
            \node (rect) [fill=redSquare,thick, inner sep= 3pt] {}}}
            
\definecolor{blueSquare}{RGB}{8, 174, 255}
\newcommand*\blueSquare[1]{\tikz[baseline=(char.base)]{
            \node (rect) [fill=blueSquare,thick, inner sep= 3pt] {}}}

\newcommand*{\rom}[1]{\expandafter\@slowromancap\romannumeral #1@}

\newcommand\alexout{\bgroup\markoverwith{\textcolor{red}{\rule[0.5ex]{2pt}{0.4pt}}}\ULon}

\title{\toolName: A Geometrical Representation for Quantum State Visualization}
\author[1125]{}
\author[Shaolun Ruan et al.]
{\parbox{\textwidth}{\centering 
        Shaolun Ruan$^{1}$,
        Ribo Yuan$^{2,1}$,
        Qiang Guan$^{3}$,
        Yanna Lin$^{4,1}$,
        Ying Mao$^{5}$,
        Weiwen Jiang$^{6}$,
        Zhepeng Wang$^{6}$,
        Wei Xu$^{7}$,
        and Yong Wang$^{1}$\thanks{Yong Wang is the corresponding author.}
        }
        \\
{\parbox{\textwidth}{\centering 
        $^1$School of Computing and Information System, Singapore Management University, Singapore\\
        $^2$Department of Computer and Information Sciences, University of Delaware, United States\\
        $^3$Department of Computer Science, Kent State University, United States\\
        $^4$Department of Computer Science and Engineering, The Hong Kong University of Science and Technology, Hong Kong, China\\
        $^5$Computer and Information Science Department, Fordham University, United States\\
        $^6$Electrical and Computer Engineering, George Mason University, United States\\
        $^7$Computational Science Initiative, Brookhaven National Laboratory, United States\\
       }
}
}
\begin{document}

% \teaser{
%  \includegraphics[width=\linewidth]{eg_new}
%  \centering
%   \caption{New EG Logo}
% \label{fig:teaser}
% }

\maketitle
%-------------------------------------------------------------------------
\begin{abstract}

Visualizations have played a crucial role in helping quantum computing users explore quantum states in various quantum computing applications. Among them, Bloch Sphere is the widely-used visualization for showing quantum states, which leverages angles to represent quantum amplitudes. However, it cannot support the visualization of quantum entanglement and superposition, the two essential properties of quantum computing. To address this issue, we propose \toolName, a novel visualization for quantum state representation. By explicitly correlating 2D geometric shapes based on the math foundation of quantum computing characteristics, \toolName\ effectively represents quantum amplitudes of both the single qubit and two qubits for quantum entanglement. Also, we use multiple coordinated semicircles to naturally encode probability distribution, making the quantum superposition intuitive to analyze. We conducted two well-designed case studies and an in-depth expert interview to evaluate the usefulness and effectiveness of \toolName. The result shows that \toolName\ can effectively facilitate the exploration of quantum states for the single qubit and two qubits.
%-------------------------------------------------------------------------
%  ACM CCS 1998
%  (see https://www.acm.org/publications/computing-classification-system/1998)
% \begin{classification} % according to https://www.acm.org/publications/computing-classification-system/1998
% \CCScat{Computer Graphics}{I.3.3}{Picture/Image Generation}{Line and curve generation}
% \end{classification}
%-------------------------------------------------------------------------
%  ACM CCS 2012
   % (see https://www.acm.org/publications/class-2012)
%The tool at \url{http://dl.acm.org/ccs.cfm} can be used to generate
% CCS codes.
%Example:


\end{abstract}

\section{Introduction}

% \yong{Need to re-write it.}

Quantum computing has been undergoing impressive development in recent years
% over classical computing
~\cite{arute2019quantum,moller2017impact}. 
An increasing number of applications have been proven possible to achieve quantum speedups
%the power of quantum computing, 
such as optimization problems, machine learning, cryptography, and finance~\cite{hassija2020present}.
% An increasing number of quantum algorithms have been developed and shown a significant speedup
% over their best-known classical counterparts,
% for example, \textit{Shor's} algorithm for integer factorization~\cite{shor1999polynomial} and \textit{Grover's} algorithm for unstructured data searching~\cite{grover1996fast,kwiat2000grover}.
Meanwhile, along with the evolution of real quantum computers provided by many quantum vendors such as IBM, Rigetti, Honeywell, and IonQ~\cite{saki2021survey}, the last decade represents a significant milestone in the field of quantum computing~\cite{corcoles2019challenges, tacchino2020quantum, gomes2018quantum}.
% Thus, there is an expectation that near-term quantum devices will soon support practical applications.

For the implementation of quantum computing, a quantum bit (\textit{a.k.a.} \textit{qubit}) is the basic unit of any quantum program.
Generally, the state of a single qubit or multiple qubits is called \textit{quantum state}, such as state $\ket{0}$ for a single-qubit state and state $\ket{01}$ for a two-qubit system.
Compared to classical computing, today's quantum computing is driven by two basic quantum-specific properties called \textit{quantum superposition} and \textit{quantum entanglement}~\cite{rieffel2011quantum, tannu2019not}, which makes it possible to achieve the power of the quantum advantages~\cite{steane1998quantum}.
Specifically, quantum superposition (\textit{a.k.a., superposition}) indicates the unknown quantum states before measurement. Upon measurement of the qubit, it collapses to either the $\ket{0}$ or $\ket{1}$ state according to the deterministic probabilities of all possible states~\cite{gokhale2020optimized}.
% (\textit{e.g.}, $Pr(\ket{0})=0.6$ and $Pr(\ket{1})=0.4$).
For example, the probability of the measured result 0 is 0.4, while that of 1 is 0.6.
Also, quantum entanglement (\textit{a.k.a. entanglement}) supports the correlations between multiple single qubits such that manipulating one qubit can impact the state of the other qubit(s). Among them, two-qubit entanglement plays a critical role in many popular applications (\textit{e.g.}, Grover's Algorithm~\cite{hayward2008quantum}, Deutsch Algorithm~\cite{gulde2003implementation}, and Quantum Machine Learning ~\cite{SamMLSys22, KerstinNature2022}), especially in the near-term Noisy Intermediate-Scale Quantum~\cite{bharti2021noisy} (NISQ) era where the qubit number is severely limited.
% The aforementioned two properties (\textit{i.e.}, superposition and entanglement) are the fundamental ingredients to .

To effectively study and execute quantum programs, observing the quantum state is a basic requirement. A \modify{visualization technique} named \textit{Bloch Sphere}~\cite{bloch1946nuclear} was proposed to meet this need. Bloch Sphere leverages a point on the unit sphere to represent the \textit{amplitudes} of a pure single-qubit state. 
Due to the intuitiveness, Bloch Sphere is widely-accepted in the quantum computing community to visually show the quantum state of a single qubit ~\cite{wie2014bloch,131387, havel2004bloch} and has been incorporated into many popular toolkits of quantum computing, such as Qiskit by IBM Quantum~\cite{ibmq}. 
Despite its prevalence, several issues still exist when using Bloch Sphere.
First, 
% as mentioned earlier, entanglement is critical to the power of quantum computing.
it is impossible to use Bloch Sphere 
%to perform entanglement using multiple qubits, not even for two qubits
to visualize quantum states of more than one qubit~\cite{bardin2021microwaves, wie2014bloch}. 
Once two qubits are entangled, the visualization approach Bloch Sphere breaks down because the quantum state cannot be condensed into three dimensions in the same form~\cite{zable2020investigating}.
Second, although Bloch Sphere can visualize quantum states, it cannot support an intuitive view of the probability of all basis states. 
However, the probability is a critical property of quantum states as it is used to directly reflect superposition~\cite{wie2020two}.
Third, three-dimension visualizations, like Bloch Sphere, have been proven to perform worse than two-dimension counterparts when conducting precise measurements~\cite{tory2005visualization, FA20153D2D}.
% Third, \textit{Bloch Sphere} is a three-dimensional visualization that performs worse than its two-dimensional counterparts
%when performing precise measurements
The prior study has found that the three-dimensional visualization of the \textit{Bloch Sphere} introduces the possibility of visual occlusion~\cite{williams2021qcvis}, which makes the observation of quantum states rather inaccurate.

% the  three above issues significantly hinder the performance and accuracy of quantum state visualization.

Thus, a novel 2D visualization approach is urgently needed to address the above issues for all quantum computing users (\textit{e.g.}, novices and experts).
% However, there are challenges that exist to propose such a visualization, which are mainly from two perspectives: \textit{individual properties of quantum states}, and \textit{complex construction of quantum entanglement}.
It is a challenging task to visualize quantum states and the challenges mainly come from two aspects:
\textit{individual properties of quantum states}, and \textit{complex construction of quantum entanglement}.
First, according to our co-design process with domain experts, 
% it is essential to lower quantum computing users' learning curve and make them learn and use the visualization smoothly.
it is essential to visually identify the correlation among all quantum state components (\textit{e.g.}, the amplitudes of the state vectors and probabilities of quantum states) other than representing the above components individually.
Because this can make users aware of the hidden logic of the evolution of quantum states.
% Using the original quantum computing characteristics to 
% visually correlate all quantum state components (\textit{e.g.}, the amplitudes of the state vectors and probabilities of quantum states) 
% other than representing the above components individually 
% can be significant for domain users.
However, how to seamlessly link the individual quantum state components by appropriate visual channels is not a trivial task.
% On the one hand, it is difficult to encode state vectors and probability distributions simultaneously using a set of visually correlated visual elements. Building upon the quantum theorems of quantum states, it is crucial to highlight the relationship between the state vectors and probability distributions, which makes it more easy-to-understand and interpretable for domain users in quantum computing.
% On the other hand,
% \yong{You are always using ``one the one hand, on the other hand'' in the wrong way.}
Second,
how to visualize the entangled quantum states remains a challenging task.
Given that the two-qubit entanglement state is a specific type of two-qubit state, the two-qubit state representation is still non-transparent for quantum computing researchers and developers since the two-qubit state is not a simple accumulation of multiple single-qubit states.
Meanwhile, how to utilize the same scalable form of visualization to represent the entangled states of both single qubits and multiple qubits would be even more difficult.

To address the above issues, we propose \toolName, a novel \underline{\textbf{V}}isual d\underline{\textbf{E}}sig\underline{\textbf{N}} for quant\underline{\textbf{U}}m state repre\underline{\textbf{S}}entation. 
\toolName\ supports quantum state representation for not only single-qubit states but also two-qubit states, which is the basis for two-qubit entanglement representation. Meanwhile, \toolName\ can inform users of the probability for the quantum superposition without any manual calculation of the probability, leading to more efficient and smooth analysis of arbitrary quantum states.
Following a user-centered design process~\cite{munzner2009nested},
% and a method~\cite{lee2022collabally} of informing the design, 
we work closely with five domain experts in quantum computing. A co-design process is conducted to iteratively derive design requirements for observing quantum states, which also guide our subsequent visual design.
\toolName\ mainly consists of two coordinated visual components: a set of right triangles to visualize state vectors, and the circumscribed semicircles of these right triangles explicitly reflect the probability distribution of the corresponding quantum states. 
To evaluate the usefulness and effectiveness of \toolName, we present two case studies on single-qubit (\textit{i.e.}, Quantum Neural Network) and two-qubit (\textit{i.e.}, \textit{Grover's algorithm}). Moreover, we conduct an in-depth interview with 14 domain experts in quantum computing. The results show that \toolName\ can effectively facilitate the visual analysis of various quantum programs.
The major contributions of this paper can be summarized as follows:

\begin{itemize}
    \item We co-design with five domain experts to identify and summarize the design requirements for visually analyzing the quantum states, pinpointing the common challenges when performing visual analysis of quantum states.
    % qua by quantum computing users.
    
    \item We present a novel visualization, \toolName, to tackle both quantum superposition and quantum entanglement representation challenges for both single-qubit and two-qubit scenarios with multiple visually correlated 2D geometrical shapes.
    
    \item We conduct two case studies and user interviews with 14 domain experts to validate the effectiveness and usefulness of \toolName.
\end{itemize}

\section{Background}

In this section, we introduce a set of relevant concepts in quantum computing,
% that are  to our work, 
including the building block of quantum computing, the properties of the qubit, and the quantum state of the qubit. 

\def\Zero{
\begin{bmatrix}
    1 \\
    0 \\
\end{bmatrix}}

\def\One{
\begin{bmatrix}
    0 \\
    1 \\
\end{bmatrix}}

\def\sv{
\begin{bmatrix}
    \alpha \\
    \beta \\
\end{bmatrix}}

\def\svcomplex{
\begin{bmatrix}
    a+bi \\
    c+di \\
\end{bmatrix}}

\subsection{Building Block of Quantum Computing}

\textbf{Qubit},
the quantum version of the classic bit, is the basic unit in quantum computing. Similar to a classical bit, there are two computational basis states called state 0 and state 1 for a qubit~\cite{ciaran2021qc4c}. 
However, 
% in contrast to a classical bit, which is either in state 
% 0 or state 1, 
a qubit can also be in an arbitrary linear superposition of state 0 and state 1~\cite{ciaran2021qc4c, rieffel2000introQCnonP}, which is well-known as quantum superposition.
Mathematically, one can represent a qubit using the form of a state vector~\cite{rieffel2000introQCnonP}.

\textbf{Quantum gate},
just like the manipulation of classical bits using classical logic instructions such as \textit{OR} and \textit{AND}, it is applied to qubits to change their quantum states. 
% and the quantum states of qubits change depending on which gate is applied. 
% In Bloch Sphere representation, gates provide instructions for rotating the qubit’s vector around the sphere~\cite{ciaran2021qc4c}. 
Mathematically, quantum gates are represented as operation matrices, acting on single qubit or multiple qubits.
Operations of quantum gates are equivalent to the dot products with the state vector of qubits.
% \yong{Applying xxx to qubits? Please double check the verb.}
% that act on qubits using matrix multiplication.
%Quantum Gate are represented by matrix. Single quantum gate like H, Multiple Quantum gate like CNOT.
%Pics of gate

\textbf{Quantum circuit},
similar to the classical circuit, is the implementation of the quantum program for execution.
A quantum circuit consists of a set of quantum gates, acting on one or multiple fixed qubits.
% in a quantum computer or quantum simulator.
As shown in Figure \ref{fig:case1} and \ref{fig:case2}, a quantum state will be initialized from the start of the quantum circuit and manipulated by quantum gates designed in the quantum circuit. 
% The final execution result is retrieved by measuring each quantum state.

\subsection{Properties of Qubit}
\label{qubit_prop}
\textbf{Superposition}
indicates that a qubit can not only be in one of the computational basis states $\ket{0}$ or $\ket{1}$, but also in a linear superposition of this two states~\cite {nara1999QC4B}.
% As mentioned in Section 3.1, compared with classical bits, the value can only be either 0 or 1, 
Thus, the quantum state $\ket{\psi}$ of a qubit is described by $\alpha\ket{0} + \beta\ket{1}$, where the complex numbers $\alpha$ and $\beta$ are called \textit{amplitudes} such that $|\alpha|^2 + |\beta|^2 = 1$~\cite{rieffel2000introQCnonP}. 
Meanwhile, 
% Once such superposition is measured with respect to the basis state {$\ket{0}$ and $\ket{1}$}, 
the probability of measuring $\ket{0}$ is $|\alpha|^2$ and the probability of $\ket{1}$ is $|\beta|^2$
~\cite{ciaran2021qc4c, rieffel2000introQCnonP,nara1999QC4B}.
% The real power of quantum computing derives from the exponentially increasing state space of multiple qubits,
as a quantum system with $n$ qubits can generate a linear superposition of $2^n$ basis states simultaneously~\cite{rieffel2000introQCnonP, Hey1999QuantumCA}.
% : since the quantum state of a single qubit can be in a linear superposition of $\ket{0}$ and $\ket{1}$, a quantum system with $n$ qubits can generate a linear superposition of $2^n$ basis states~\cite{rieffel2000introQCnonP, Hey1999QuantumCA}. 
% Thus, creating such a superposition is the critical properties of achieving the quantum advantage~\cite{Hey1999QuantumCA}.

% that gives quantum parallel processing its power~\cite{Hey1999QuantumCA}.

% Example.
% It is noteworthy that even though a qubit can have infinitely many superposition states before measurement, measuring a qubit collapses its superposition state into one of two possibilities~\cite{ciaran2021qc4c, rieffel2000introQCnonP}. That is, if the measurement of $\ket{\psi} = \alpha\ket{0} + \beta\ket{1}$ results in $\ket{0}$, then the state changes to $\ket{0}$, and a second measurement with respect to the same basis will return $\ket{0}$ with probability 1~\cite{rieffel2000introQCnonP}. Thus, unless the original state happens to be one of the basis vectors, the measurement will change that state, and it is impossible to determine what the original state was~\cite{rieffel2000introQCnonP}.

\textbf{Entanglement}
is an essential feature that differentiates qubits and classical bits. 
Specifically,
% In quantum computing, 
when two or more qubits are entangled, their quantum states are coupled with each other,
%they will exist in a single quantum state. 
so that changing the quantum state of any one qubit will instantaneously change the other qubit's quantum state in a predictable way~\cite{rieffel2000introQCnonP}.
% Instead, an easy test to determine if a system is entangled or not is to check if measuring the value of one qubit changes the probability distribution of the other qubit~\cite{ciaran2021qc4c}. 
% For instance, the quantum state $\ket{\psi} = \dfrac{1}{\sqrt{2}}\ket{00} + \dfrac{1}{\sqrt{2}}\ket{11}$ is entangled since the probability that the first qubit is measured to be $\ket{0}$ is $1/2$ if the second qubit has not been measured. 
Note that the entanglement operation requires more than one qubit, making it critically important to analyze the quantum states of multiple qubits instead of a single qubit.
% Note that the two-qubit entangle is a specific scenario of two-qubit state, where the two qubits are entangled.
% Thus, the approach of two-qubit state representation is of great value for solving the two-qubit entanglement representation.
% However, if the second qubit has been measured, the probability that the first qubit is measured as $\ket{0}$ is 1 or 0, if the second qubit was measured as $\ket{0}$ or $\ket{1}$, respectively. In this case, the result of measuring the first qubit depends on the measurement of the second qubit. 
% In contrast, in the case of $\ket{\psi} = \dfrac{1}{\sqrt{2}}\ket{00} + \dfrac{1}{\sqrt{2}}\ket{01}$, two qubits are not entangled since any measurement of the first qubit will yield $\ket{0}$ regardless of whether the second qubit was measured. Similarly, the second qubit has a fifty-fifty chance of being measured as $\ket{0}$ regardless of whether the first qubit was measured or not~\cite{rieffel2000introQCnonP}. Entanglement happens even if qubits are far away from each other. When two or more qubits are entangled, the quantum state can not be described in terms of the state of each of its components (qubits) separately. 

\subsection{Quantum State of Qubit}
\label{sec:3.3}
In quantum computing, a quantum state is a mathematical entity that provides a probability distribution of different basis states.
% on a single qubit or multiple qubits. 
For clarity, we start with a single-qubit state, and the case of a two-qubit state will be derived from these results.
In Section \ref{sec:venus}, we will illustrate how we apply the following quantum computing characteristics and encode them with a variety of 2D geometric shapes to form our visual design.

\textbf{Single-qubit state} represents the quantum state for a single qubit.
% From Section 3.1 and 3.2,
Recall that
the quantum state of a qubit can be a superposition of basis states {$\ket{0}$ and $\ket{1}$}, thus the quantum state $\ket{\psi}$ can be expressed as $\ket{\psi} = \alpha\ket{0} + \beta\ket{1} = \alpha\Zero + \beta\One = \sv$, where the amplitudes $\alpha$ and $\beta$ satisfy:
% are complex numbers
% , and $|\alpha|^2 + |\beta|^2 = 1$. 
% and can be calculated as follows:

\begin{equation}
\label{equation:1}
\alpha=a+b\cdot i, \beta=c+d\cdot i,
\end{equation}

where $i$ is the imaginary unit, and $a$, $b$, $c$, and $d$ are real numbers.
% Thus, the quantum state of a single qubit can be expressed as $\ket{\psi} = \sv = \svcomplex$. 
Based on the quantum 
% computing 
theory, the probability of a measured quantum state (\textit{e.g.}, $\ket{0}$) satisfies

\begin{equation}
\label{equation:2_1}
Pr(\ket{0}) = |\alpha|^2 = |a|^2 + |b|^2.
\end{equation}

% By definition, a complex number is a number of the form $a + bi$, where $a$ and $b$ are real numbers, and $i$ is an indeterminate satisfying $i^2=-1$. 
Meanwhile, since the amplitudes satisfy a normalization constraint,\textit{ i.e.}, the sum of the probabilities of the two quantum states for single qubits (\textit{i.e.}, $\ket{0}$ and $\ket{1}$) consistently equals 1, 
thus applying Equation \ref{equation:2_1}  yields

\begin{equation}
\label{equation:4}
|a|^2+|b|^2+|c|^2+|d|^2 = 1.
\end{equation}

% $a$, $b$, $c$ and $d$ satisfy 
% $a^2+b^2+c^2+d^2=1$.

\textbf{Two-qubit state} is the quantum states executing on a pair of qubits,
\modify{which can be calculated by the tensor product of two single-qubit states, \textit{e.g.}, $\ket{00} = \ket{0} \otimes \ket{0}$}.
% \yong{Pls check my comments on Overleaf Review.}
Meanwhile, any two qubits can be in the state $\ket{\psi} = \alpha\ket{00} + \beta\ket{01} + \gamma\ket{10} + \delta\ket{11}$, where the amplitudes $\alpha$, $\beta$, $\gamma$, and $\delta$ satisfy:

% For two-qubit entanglement, the state vector consists of four complex number amplitudes (\textit{i.e.}, $\alpha-$ $\beta-$ $\gamma-$ $\delta-$components), which can be represented as follows:

\begin{equation}
\label{equation:5}
\begin{aligned}
\alpha = a + b \cdot i, \beta = c + d \cdot i,\\
\gamma = e + f \cdot i, \delta = g + h \cdot i.
%  &\quad
\end{aligned}
\end{equation}

Similar to single-qubit states, since the probabilities of all possible qubits equal 1, the four amplitudes satisfy  $|\alpha|^2 + |\beta|^2 + |\gamma|^2 + |\delta|^2 = 1$. 
By applying Equation \ref{equation:5} to the above constraint,
we have
% By applying Equation \ref{equation:5} to Equation \ref{equation:6}, we can calculate the lengths of all line segments for state vector display as follows:

\begin{equation}
\label{equation:7}
|a|^2+|b|^2+|c|^2+|d|^2+|e|^2+|f|^2+|g|^2+|h|^2 = 1.
\end{equation}

% \begin{figure}[t]
% \centering 
% \includegraphics[width=0.7\linewidth]{figures/1_.pdf}
% \caption{A widely-used visualization called Bloch Sphere for quantum state representation.
% % (B) The workflow of the co-design process with domain experts. The steps connected by the green line indicate the stage of the preliminary interview for the initial prototype, while those steps connected by the blue line represent the iterative tuning process in the stage of the expert test.
% }
% \label{fig:1}
% \end{figure}
\section{Related Work}

Our work is relevant to prior studies on graphical tools in quantum computing and the visual design of quantum states.

\subsection{Graphical Tools in Quantum Computing}

Researchers have attempted to visualize quantum circuits using graphical interfaces,
% Much prior work focused on, 
including quantum circuit implementation
% ~\cite{gheorghiu2018quantum,paykin2017qwire}
and building graphical simulators for quantum circuits.
% ~\cite{kelly2018simulating,zulehner2018advanced, moran2016quintuple}.
Specifically, Paykin et al.~\cite{paykin2017qwire} presented an interface to manipulate quantum circuits using a classical host approach. 
Zulehner et al.~\cite{zulehner2018advanced} proposed a new graph-based approach for quantum simulators.
Much prior work has also been studied to interpret quantum algorithms and workflow using a graphical approach.
% ~\cite{tao2017shorvis,karafyllidis2003visualization, ruan2022vacsen,miller2021graphstatevis,williams2021qcvis,lin2018quflow}.
Ruan et al.~\cite{ruan2022vacsen} introduced a real-time visualization system for noise awareness in quantum computers and compiled circuits.
Tao et al.~\cite{tao2017shorvis} propose a tool to assist users in understanding Shor's algorithms using a graphical interface.
% Karafyllidis et al.~\cite{karafyllidis2003visualization} also leveraged the visualization approach to help novices understand QFT (\textit{i.e.}, Quantum Fourier Transform) algorithm.
In addition, 
% cloud platforms support online quantum circuit implementation~\cite{quirk, qcircuits, ibCirqmq}.
% For example, 
\textit{Quirk}~\cite{quirk} also provides a graphical interface to make users aware of the quantum circuit's behavior. 
% Many large IT companies also provide cloud platforms that allow users to implement quantum circuits online, such as IBM Quantum~\cite{ibmq}.
The aforementioned interfaces focused on improving the interpretability of quantum circuits using 
% a graphical interface, 
% they mainly utilize 
multiple fundamental visualizations indivisually (\textit{e.g.}, bar charts, circles in \textit{Quirk}~\cite{quirk}). 
However, this type of visualization cannot intuitively visualize the relationship between properties in quantum computing, making the users feel struggle to understand the changes happening in quantum states.
Our work addressed this challenge by introducing a visual solution, assisting users in quickly understanding single-qubit and two-qubit states thoroughly.

\subsection{Visual Design of Quantum States}

Many prior studies have focused on visualizing quantum states using 
% a variety of methods, which can be categorized into
3D and 2D visualizations.
For 3D representation, 
\modify{a} Block Sphere~\cite{bloch1946nuclear} visualizes single-qubit states based on a 3D geometrical representation and is still the widely-used visual representation till now~\cite{131387, havel2004bloch}.
Some prior work focused on extending \modify{Bloch Spheres}.
For example,
% ~\cite{makela2010n,altepeter2009multiple,wie2020two}.
Altepeter et al.~\cite{altepeter2009multiple} extended Bloch Sphere using the remote-state preparation protocol.
% Boyer et al.~\cite{boyer2017geometry} provided a geometrical analysis of entanglement and separability for all mixed states.
IBM also provides a Bloch Sphere-like design called Q-Sphere~\cite{ibmq} to represent multiple states in a single sphere.
% Tamate et al.~\cite{tamate2011bloch} proposed a way to represent three-vertex geometric phases for multiple state systems on the Bloch sphere.
In addition, many researchers have studied how to represent quantum states using 2D shapes other than extending Bloch Sphere.
For example,
% ~\cite{galambos2012visualizing, lopez2018geometry, avanesov2019unitary, chernega2017triangle}.
Galambos et al.~\cite{galambos2012visualizing} utilized fractal representation to visualize the multi-qubit qubit systems.
Chernega et al.~\cite{chernega2017triangle} mapped the density matrix of the qubit onto the vertices of a triangle. 
% Avanesov et al.~\cite{avanesov2019unitary} offered a method to visualize the quantum channel’s maps of qubit states.
While the prior studies provided different ways to portray the quantum states, 
none of the prior work focused on the representation of superposition, which reflects the probability of measuring each quantum state.
We aim to visually correlate multiple properties with the probability of each quantum state, making the users understand the probability more comprehensively.
%without any manual calculation.

\section{Informing The Design}

% In this section, we discuss the stages of our co-design process and the design requirements we summarized during the process.
In this section, we introduce our co-design process and the derived design requirements from it, which will inform our subsequent development of the visual design.

\subsection{Co-design Process}

The primary goal of our co-design process is to collect the task abstractions faced by quantum computing users.
% which are used to derive the design requirements of our visual design.
Thus, following the guideline of design study~\cite{sedlmair2012design}, we worked closely with five domain experts for over five months.
First, we conducted preliminary interviews with five quantum computing experts, where we sought to understand the practical challenges and difficulties.
Second, over the next four months, we revised our initial prototype iteratively according to the feedback collected from the expert test.
% The co-design process is shown in Figure \ref{fig:1}\component{B}.
% We present the co-design process and summarize the design requirements which are used to inform our visual design, \toolName.

\textbf{Participants.}
We invited five domain experts \textbf{P1-5} (5 males, $age_{mean}= 32.13$, $age_{sd}= 4.71$) to participate in our co-design process.
Specifically, \textbf{P1} is a research scientist from Pacific Northwest National Laboratory in the U.S., \textbf{P2-5} are either professors or post-doc researchers from three different universities in the U.S. 
Among them, the research direction of \textbf{P1-2} and \textbf{P5} is Quantum Machine Learning, while \textbf{P3-4} are working on Quantum Systems.
All the domain experts have an average of 6.3 years of research and development experience in quantum computing.

\textbf{Preliminary Interview.}
Following the methodology proposed by Sedlmair et al.~\cite{sedlmair2012design},
we began the preliminary interviews by performing one-on-one, semi-structured, hour-long interviews with Group 1 (\textbf{P1-3}), to collect current challenges the participants have when working on quantum computing.
First, each participant was asked to describe the major issues and challenges they faced using the most popular visualization, \textit{i.e.}, \textit{Bloch Sphere}.
Note that each participant in Group 1 used \textit{Bloch Sphere} as the common tool in their daily routine tasks.
Through this session, we collected a set of high-level requirements about the limitations of the \textit{Bloch Sphere} representation.
The aforementioned tasks last about 25 minutes.
Then, we further asked every participant to describe the future requirements in a think-aloud manner, including those that can facilitate the visual analysis of quantum states.
We summarized these initial requirements and expectations during this session.
This session lasted about 30 minutes.
We video-recorded and took notes for each interview and discussion.
All requirements collected in this round were used to inform our initial prototype.

\textbf{Expert Test.}
Over the next four months, we focused on iteratively testing the features of our initial prototype with Group 2 (\textbf{P4-5}).
We designed and implemented the visual design according to the qualitative requirements collected from Group 1
% in the previous session
and released it as an online interface, which allows Group 2 to access and use it.
Specifically, we began the expert test by briefly introducing the session's purpose and describing the initial prototype's functionality.
Each participant was encouraged to use our design when performing their daily domain tasks in quantum computing.
We collected their feedback by holding iterative meetings with the participants every two weeks, which were open and loosely structured to ensure they could express anything they thought of. 
We further tuned the several design alternatives (as shown in Fig. \ref{fig:32}) accordingly to guarantee our design meets all these practical needs.

\subsection{Design Requirements}
\label{sec:design_requirements}

We summarized all design requirements from the co-design process.
% , including both the initial feedback from Group 1 and subsequent requirements from Group 2.
% summarized
% from the stage of during the stage of the expert test. 
We reported six refined requirements and categorized them into \textbf{functionality} and \textbf{usability}.

For the functionality, participants reported three major requirements to facilitate the observation of arbitrary quantum states:
% which need to be enhanced from the following aspects:

\begin{itemize}
  
    \item[\textbf{R1}] \hspace{7pt}\textbf{Visualize two-qubit states in addition to one-qubit states.}
    All participants (\textbf{P1-5}) reported a strong need to support the visualization of entanglement for two qubits. Thus, the basis for two-qubit entanglement representation is how to represent a two-qubit state.
    % entanglement of qubits, especially for the two-qubit state visualization. 
    \textbf{P3} also suggested that it would be helpful to represent single-qubit and two-qubit states in the same set of visualization other than using two different forms of approaches.

    \item[\textbf{R2}] \hspace{7pt} \textbf{Provide an intuitive representation for the probability distribution of different states.}
    All participants (\textbf{P1-5}) suggested that visually reflecting the probability distribution is essential for the quantum state exploration as the probabilities of different states explicitly reveal the \textit{superposition}, which is the fundamental ingredient of quantum computing. 
    % Thus, they believe it is crucial to
    Also, they preferred a visualization that can naturally show the probabilities with visual elements without any time-consuming manual calculation.

    \item[\textbf{R3}] \hspace{7pt} \textbf{Inform users of the state vector of each quantum state.}
    Four participants (\textbf{P1-3, P5}) confirmed that showing the state vector (\textit{e.g.}, amplitude $\alpha$ and $\beta$) as-is will greatly benefit quantum computing users. 
    % \textbf{P1} also mentioned that showing the essential individual element of each state in a state vector will benefit all users.
    \textbf{P5} also reported that forming the visualization with states vectors is better for any quantum computing users due to every visual element's intuitiveness.

\end{itemize}

For usability, participants focused on 
% the needs on 
how to make full use of the visualization for domain users, which was summarized as follows:

\begin{figure*}[t]
\centering 
\includegraphics[width=0.9\linewidth]{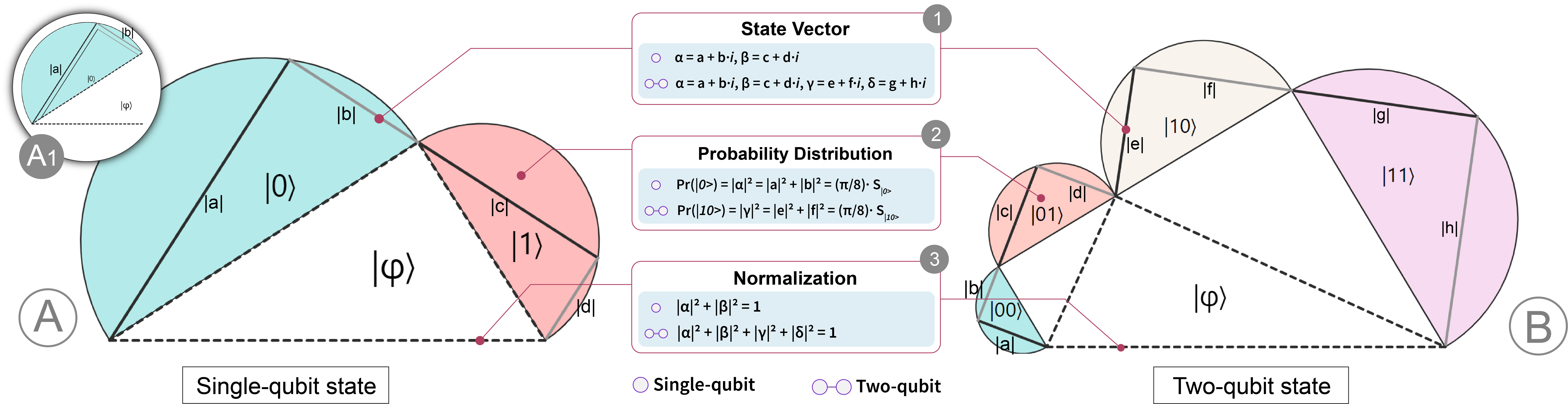}
\caption{The visual design of \toolName\, which supports single-qubit (A) and two-qubit (B) state representation based on the same visualization form. 
\textbf{Line segments} visualize the state vector, where the black line denotes the real part, and the grey line denotes the imaginary part based on Equation \ref{equation:1}.
\textbf{Semicircles}'s area indicates the probability of measuring the corresponding state based on Equation \ref{equation:2_1}.
\textbf{Triangle base}'s length consistently equals to 1, because it encodes based on the constraint of normalization (\textit{e.g.}, Equation \ref{equation:4}).
}
\label{fig:2}
\end{figure*}

\begin{itemize}
  
    \item[\textbf{R4}] \hspace{7pt} \textbf{Visually correlate elements in quantum states.}
    % Three participants (\textbf{P1-2}, \textbf{P5}) emphasized that it is significant to make the visual design easy-to-understand for quantum computing users.
    Three participants (\textbf{P1-2}, \textbf{P5}) emphasized that it is significant to represent the state probabilities naturally by the basic element (\textit{e.g.}, state vectors) other than an additional individual visual channel,
    since visually correlating different individual elements by the quantum mechanics theory makes the evolution of quantum states more easy-to-understand.
    % visually correlating different individual elements for a given quantum state (\textit{i.e.}, state vectors and the corresponding probabilities) and 
    % encoding the quantum mechanics theory that the quantum computing users are familiar with into the visual elements will be helpful for domain users.

    \item[\textbf{R5}] \hspace{7pt} \textbf{Display all visual elements with 2D shapes.}
    Three participants (\textbf{P3-5}) confirmed that they prefer a 2D representation to 3D of a quantum state.
    For example, the \textit{Bloch Sphere} - the visualization tool that quantum computing users use most - is a 3D sphere. They commented that it is tough to locate the point's location and read the rotation angles accurately, even though the interaction of dragging is supported in some tools.

    \item[\textbf{R6}] \hspace{7pt}\textbf{Make the visualization tool accessible for all quantum computing users.}
    Even though four participants (\textbf{P1-2}, \textbf{P4-5}) gave positive feedback for the useful prototype interface to assist their routine tasks during the stage of expert tests, they also emphasized the necessity of making it publicly-available to benefit all quantum computing users. 
    Considering the power and popularity of the web-based cloud quantum computing platforms (\textit{e.g.}, IBM Quantum), an online web interface will be a good choice.

\end{itemize}

\section{\toolName}
\label{sec:venus}

Drawing from the iterative co-design process, we designed \toolName, a novel visualization approach that facilitates the observation of the single-qubit and two-qubit quantum states.
% Users can easily access the \toolName\ interface at 
To enhance accessibility, we implemented a web-based graphical interface to fit into routine tasks.
Users can access \toolName\ interface via
\textcolor{blue}{\url{https://venus-interface.github.io/}}.
% Design justification and a detailed introduction of the online interface are provided in Appendix A and B.
In this section, we first introduce the visual design of our visualization approach for single-qubit states and extend it to the form of two-qubit states.
% then present all design alternatives before we derive the final design at last,
% and detail the open-sourced interface of \toolName.

\subsection{Single-qubit State Representation}
\label{sec:single}

Informed by the design requirements introduced in Section \ref{sec:design_requirements}, we propose a novel visual design for single-qubit state representation. As shown in Figure \ref{fig:2}\component{A}, we utilize three right triangles and two circumscribed semicircles to represent the quantum states.

\textbf{State vector.}
According to the quantum computing theory, the state vector of a quantum state is the fundamental discipline to represent a quantum state. Thus, it is of great importance to encode the state vector in the visual design (\textbf{R3}).
Specifically, we utilize each amplitude's real and imaginary parts (\textit{e.g.}, $a$ and $b$ of amplitude $\alpha$) to indicate the state vector according to the quantum theory illustrated in Section \ref{sec:3.3}.
Based on Equation \ref{equation:1}, we utilize two line segments to depict the amplitudes of the state vector as shown in Figure \ref{fig:2}\Subcomponent{1}.
Also, we use \modify{cyan} and red right triangles to indicate the two amplitudes $\alpha$ and $\beta$ of the state vector, respectively.
We encode the absolute values of real parts (\textit{i.e.}, $a$ and $c$) and imaginary parts (\textit{i.e.}, $b$ and $d$) of each amplitude by the line segments in black and grey, respectively.
We apply the double lines, as shown in Figure \ref{fig:2}\subcomponent{A1}, to reflect the negative real and imaginary values.
Note that if the number of the imaginary part is zero, the black line segment for the real part will coincide with the diameter of the semicircle.

\textbf{Probability of the quantum states.}
According to the refined requirement, the probability of each state (\textit{e.g.}, $\ket{0}$) is significant in revealing the superposition in quantum computing (\textbf{R2}).
Based on Equation \ref{equation:2_1}, we already know that the probability of a quantum state can be calculated by the absolute value of $a$ and $b$.
Meanwhile, for \toolName, according to the geometry of the visual design, the area of the semicircle circumscribed by the right triangle indicating the state vector can be calculated as follows:

\begin{equation}
\label{equation:2}
S_{semicircle} = \frac{\pi}{8} \cdot (|a|^2 + |b|^2),
\end{equation}

where $a$ and $b$ are the real and imaginary parts of $\alpha$.
Thus, building upon the above two equations, we can represent the probability of a specific quantum state by the area of the semicircle circumscribed by the right triangle since the area of the semicircle is proportional to the probability of the quantum state, as shown in Figure \ref{fig:2}\Subcomponent{2}.
% Thus, we made it possible to encode the probability of each quantum state by the intuitive area channel of the semicircles.
% For the color scheme,  we plot the semicircles for the two quantum states in two colors blue and red for $\ket{0}$ and $\ket{1}$ respectively, .
% In the scenario of the single-qubit quantum state, there are two possible quantum states (\textit{i.e.}, $\ket{0}$ and $\ket{1}$).
Thus, it is apparent to visually analyze the two probabilities of $\ket{0}$ and $\ket{1}$ by the semicircle area based on the numerical state vectors without any manual calculation for the probability.
Also, users are allowed to visually analyze how a certain state vector (\textit{i.e.}, line segments) affects the corresponding probability.

\textbf{Correlations between all elements.}
Through the co-design process, all participants pointed out that it is significant to visually link all various elements based on the normalization constraint, because this can
% correlate all above elements (i.e., state vector and corresponding probability), 
highlight how state vector and corresponding probability affect each other (\textbf{R4}).
As shown in Figure \ref{fig:2}\Subcomponent{3}, we arrange all elements in quantum states into a systematical form of shapes.
% Specifically, given that the sum of the probabilities of all single-qubit quantum states satisfies
% \begin{equation}
% \label{equation:3}
% Pr_{\ket{0}} + Pr_{\ket{1}} = 1.
% \end{equation}
We arrange the visual channels of all elements in the quantum state into a set of right triangles with the base side length of 1 according to Equation \ref{equation:4}.
Building upon this rule, the bottom right triangle in white is used to link the other two right triangles in \modify{cyan} and red for two states (\textit{i.e.}, $\ket{0}$ and $\ket{1}$).
% we visually correlate all elements and then form the primary visual design of \toolName.

\subsection{Two-qubit State Representation}

From the co-design process, all domain experts strongly agreed that it would be much more beneficial if the visualization could support the two-qubit state representation (\textbf{R1}), which is one of the major limitations for the widely-used approach, \textit{i.e.}, Bloch Sphere.
Recall that the two-qubit state representation is the basis to make the two-qubit entanglement representation available.
We extend from the approach for single-qubit state illustrated in Section \ref{sec:single} to unveil the mask of the two-qubit quantum states.

First, according to \textbf{R3}, we intend to present the two-qubit state based on the state vector. 
As shown in Figure \ref{fig:2}\component{B},
we utilize four pairs of line segments to visualize $\alpha, \beta, \gamma$ and $\delta$ building upon Equation \ref{equation:5}.
Each pair of line segments consists of a line segment in black to represent the real part and a line segment in grey to indicate the imaginary part of the complex number amplitudes (Figure \ref{fig:2}\Subcomponent{2}).

Second, 
the probability distribution of quantum states is also supported for \toolName's two-qubit mode (\textbf{R2}).
In this case, there are four possible quantum states, \textit{i.e.}, $\ket{00}, \ket{01},\ket{10},$ and $\ket{11}$.
By applying Equation \ref{equation:2_1}, the probabilities of the above four quantum states are proportional to the area of the corresponding circumscribed semicircle.
So we encode the four two-qubit quantum states by four semicircles colored blue, red, almond, and purple. 
In this way, we convert the probability distribution for the two-qubit entanglement case by the graphical area of the four semicircles.

Third,
inspired by \textbf{R4}, we correlate all elements for entangled two-qubit states (\textit{i.e.}, four quantum states' state vectors and corresponding probabilities) by the geometry of \toolName\ intrinsically.
Compared to the single-qubit representation, we utilize three auxiliary triangles (the white triangles in Figure \ref{fig:2}\component{B}) to correlate all colored triangles for quantum state representation based on Equation \ref{equation:7}.
It is clear that the base length of the other two triangles equals the bottom triangle's side length, which is always 1.
% As the three auxiliary triangles are right triangles, the sum of the square of each line segment for state vectors, which aligns with 
% Based on this, we linked all entangled two-qubit states together in the visual design by leveraging three auxiliary triangles.

% \subsection{Implementation of \toolName}
% Before the final implementation of the visual design, we came up with several design alternatives to meet the initial requirements, whose detailed justification is provided in Appendix A. To enhance the accessibility and impact of the design, we implemented a web-based graphical interface to fit into the routine tasks of all types of users in quantum computing. The online interface is introduced in Appendix B. 
% For the color selection of \toolName, we adopted different colors with high transparency to indicate the probability distribution due to the categorical quantum states.

\begin{figure}[tbhp]
\centering 
\includegraphics[width=0.9\linewidth]{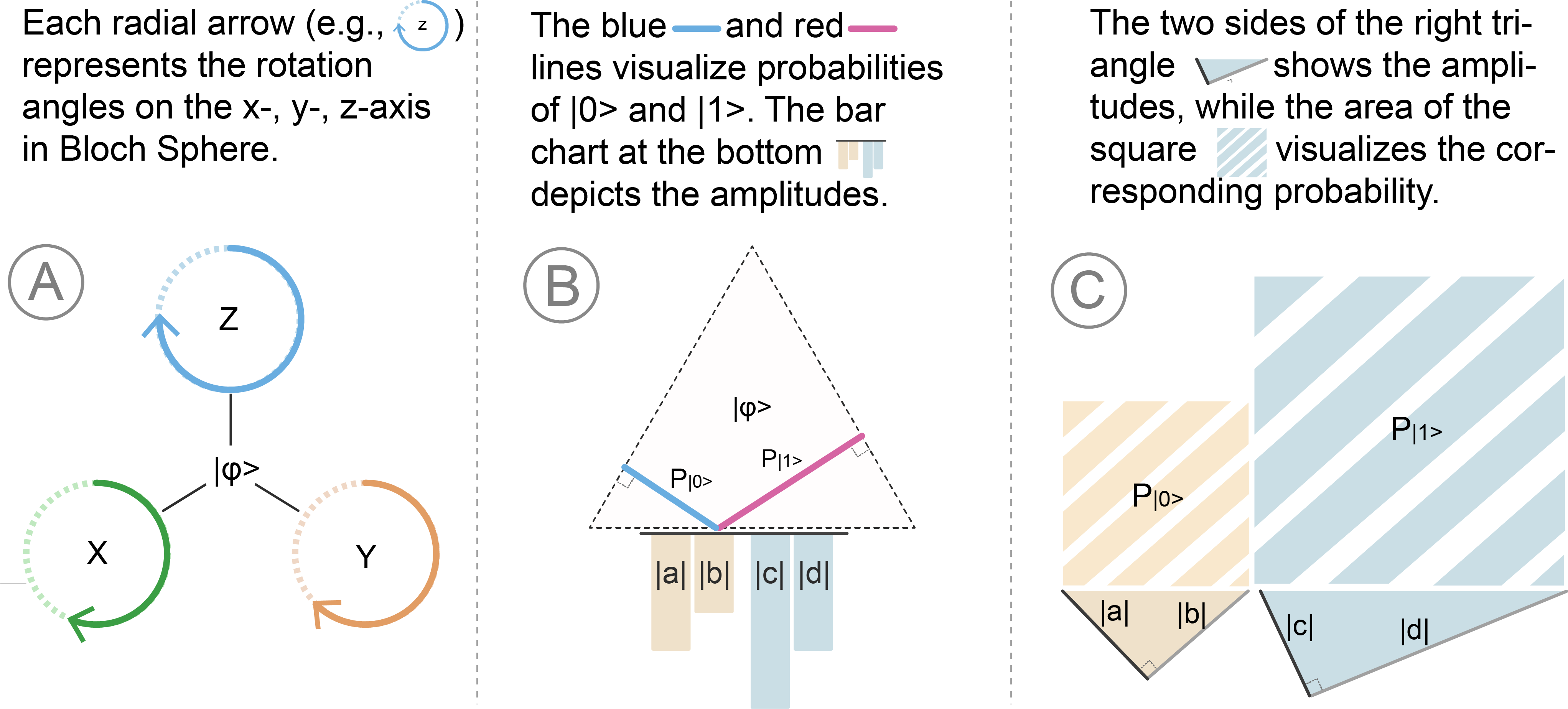}
\caption{
The design alternatives of \toolName. 
(A) The visualization simply converts the three angles of the Bloch Sphere to 2D shapes.
(B) A triangle-like design that visualizes the probabilities of two basic states (\textit{i.e.}, $\ket{0}$ and $\ket{1}$) using \textit{Viviani's theorem}.
(C) The design utilizes the square and equilateral triangle area to explicitly display the correlation between the probability and amplitudes.
% \yong{It is necessary to briefly introduce the three different designs in the caption.}
% (A) The design alternative where the three radial axes convert the original Bloch Sphere to the 2D rotation of each axis. 
% (B) The design alternative can visualize the state vector and the probability distribution by applying the geometry principle of the equilateral triangle.
% (C) The design alternative can visually depict the state vector and probability distribution using line segment length and square area, respectively.
}
\label{fig:32}
\end{figure}

\subsection{Design Justification}

\begin{figure*}[t]
\centering 
\includegraphics[width=0.9\linewidth]{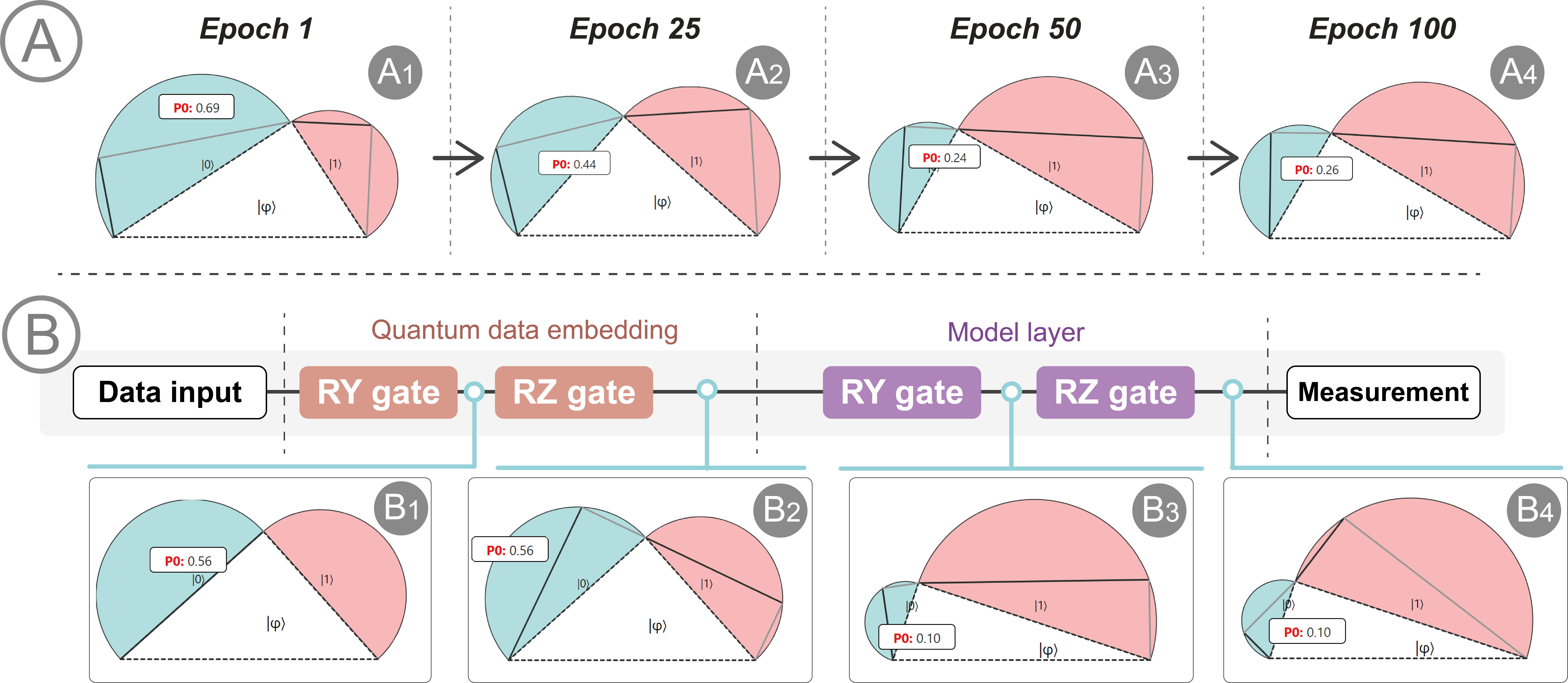}
\caption{
The case for the single-qubit quantum classifier. (A) The learning process of the quantum classifier for the Iris dataset. 
The four charts indicate the evolution of a quantum classifier from Epoch 1 to Epoch 100, where the probability of state 1 is 31\% at Epoch 1 and increased to 74\% after 100 learning epochs.
(B) The quantum states of an Iris data point in each step, which consists of two stages (\textit{i.e.}, quantum data embedding and the model layer).
`$P0$' indicates the probability the measurement result is 0.
}
\label{fig:case1}
\end{figure*}

\modify{We considered three design alternatives before we came up with the final design for quantum state representation, as shown in Figure \ref{fig:32}.}
\modify{Specifically, Figure \ref{fig:32}\component{A} shows the three axes (\textit{i.e.}, x-axis, y-axis, and z-axis) of Bloch Sphere, which simply converts the quantum states in Bloch Sphere representation to 2D shapes. However, this approach cannot display the state vector and probability distribution.}
\modify{Figure \ref{fig:32}\component{B} encodes the probabilities of the two single-qubit states (\textit{i.e.}, $\ket{0}$ and $\ket{1}$) by the length of blue and red line segments within an equilateral triangle,
% highlighted in blue and red, 
whose sum is a constant (\textit{e.g.}, the sum of two probabilities that equals 1) due to the geometrical principle of equilateral triangles, \textit{i.e.}, \textit{Viviani's theorem}~\cite{kawasaki2005proof}.
% \yong{1. Why must it be 1? It should depend on the side length of the triangle. 2. Any reference to the geometrical principle?}
}
% which applies the geometry that the sum of the distances from an arbitrary point on the base of an equilateral triangle to the two sides is a constant value. So we apply this rule to encode the two probabilities whose sum equals 1. 
\modify{However, 
% through the co-design process with domain users, they 
the domain experts pointed out in our co-design process
% argued 
that this design could not support the two-qubit state scenarios.}
\modify{Figure \ref{fig:32}\component{C} leverages two or four combinations of a square and a right triangle to visualize single-qubit and two-qubit states, respectively, where the length of two sides of triangles indicates the real and imaginary parts of each vector amplitudes.}
\modify{However, it is difficult to analyze the relationship of all quantum states and how the state vector will affect the probability. }
\modify{Thus, we further proposed the final designs (Figure \ref{fig:2}) that meet all the above requirements.}

\section{Case Study}

% In this section, we conducted two case studies to demonstrate the effectiveness of \toolName. 
We utilized two applications, \textit{i.e.}, a two-class quantum classifier and \textit{Grover's algorithm}, to evaluate \toolName\ for single-qubit and two-qubit state representation, respectively.
% The users involved in the case studies are two quantum computing experts (E4 and E9) who also attended the expert interviews in Section \ref{sec:user_study}.
The participants were asked to 
% perform the applications with the online interface and 
use a monitor with 1920 $\times$ 1080 resolution in advance.

\subsection{Case Study \uppercase\expandafter{\romannumeral 1} - Single-qubit Quantum Classifier}

E4 employed \toolName\ to explore various quantum states at the different stages of a quantum classifier.
% Specifically, E4 expected to explore a quantum neural network for two-class classification.
% Meanwhile, E4 attempted to implement the quantum classifier with the quantum simulator to avoid noise and enhance the fidelity of the measurements.
Specifically, E4 utilized Iris datasets~\cite{iris} to train the quantum classifier with a quantum simulator.
Following the design methodology~\cite{Pennylane_example}, E4 encoded a single qubit by applying the two features of the Iris dataset.
According to the network architecture proposed by Stein et al.~\cite{SamMLSys22}, E4 first implemented the classifier circuit (Figure \ref{fig:case1}) using TorchQuantum~~\cite{TorchQuantum}, which can support a flexible output for quantum states at each stage of the classification.
% Figure \ref{fig:case1} shows the circuit to execute the quantum neural network.

\textbf{Understanding the learning process of the quantum classifier.}
E4 was curious about how the learning process will act on the data points. \textit{``Generally, I faced the density matrix of the gates and qubits only, so it will be interesting to check out the evolution of the quantum states visually.''}
Keeping this in mind, E4 first selected a data point from the validation sets, whose two features were 1.4595 and 0.6797, respectively.
After the selection, E4 glanced at the quantum state output at the first epoch out of 100.
As shown in Figure  \ref{fig:case1}\subcomponent{A1}, it is clear that the probability of state 0 is slightly larger than state 1, as indicated by the area of the two semicircles.
E4 was surprised that \toolName\ shows the state probability from the amplitude values directly without any calculation
% , which is very useful and time-saving.
% ``I just type in the two complex numbers of the state vector of the quantum state, and it shows the corresponding probability of the two states at once without any calculation from me. That is very useful and helpful.''E4 commented.
E4 then noticed that the label of the data point is 1
% , which is represented by state 1 in the quantum classifier.
\textit{``Due to Label 1, I expect to see when the semicircle on the right will be larger than the left one, and how the state vector affects this convergence process.''}
So E4 evenly output the results of the data point's quantum states in subsequent three epochs (\textit{i.e.}, 25, 50, 100) to see the evolution of the quantum classifier.
As shown in Figure \ref{fig:case1}\subcomponent{A2}, the area of the semicircles indicating the probability of state 0 and 1 are almost the same at Epoch 25, and then the semicircle area of state 1 is significantly larger than that of state 0 at Epoch 50, and almost remains the same at Epoch 100.
E4 commented that it is apparent that the quantum classifier has converged around Epoch 50 for this data point.
Meanwhile, E4 noticed that this probability change was caused by the evolution of the corresponding state vector indicated by the line segments within the semicircles.
% Based on the finding, E4 then reported that The change of the two sets of line segments in states 0 and 1 have the same pattern: the real parts have increased and the imaginary parts have decreased according to the line segments.
In addition, E4 found that the increase of the real part in state 1 mainly results in the larger area of the red semicircle.
% \textit{``''}
E4 then confirmed,
\textit{``These findings provide me with a good guideline for model tuning. For example, during the parameter initialization of the rotation gates, I will choose those rotation angles which could make the absolute value of real parts larger than the imaginary part for both state 0 and 1.''} 
% \yong{The first quotation mark should be ``, NOT ''. Please check it throughout the paper.}
% After the learning process exploration, E4 summarized that
% \toolName\ provides him with an intuitive representation of the state probability distribution and naturally reveals insights into how the state vector will affect the prediction via the valuable and concise 2D shapes.
% The data points' state vectors. More importantly, it makes me naturally aware of the probability distribution and 

\textbf{Unveiling the mask of the prediction process of QNN.}
Building upon the trained quantum classifier, E4 attempted to perform the prediction process and inspect the propagation of data points.
First, E4 mentioned that despite the importance of quantum data embedding, 
it is still hard for him to understand how the original decimal data points are transformed into abstract quantum states.
% is a crucial stage to initially embed original datasets into quantum states that is a valid form for the subsequent trainable quantum gates.
% However, E4 reported that, 
% image what the embedded quantum states look like as the complex matrix cannot reflect the original decimal data points intuitively.
% Thus, E4 expected to figure out the embedded quantum states by \toolName.
\modify{E4 hinted that the whole prediction process consists of four gates, where the first two gates are used for quantum data embedding and the last two gates correspond to the trained classification model,
% for the prediction, 
as shown in Figure \ref{fig:case1}.}
Hence, E4 first randomly selected a data point and then visualized it after the first gate (\textit{i.e.}, RY gate).
\modify{E4 was surprised that the probabilities of the two states were around 0.50 (Figure  \ref{fig:case1}\subcomponent{B1})}:
% \textit{``It is interesting to me because \toolName\ provides me an intuitive image of quantum states after embedding instead of the density matrix consisting of complex numbers.'' }E4 reported.
% the only thing I encountered during the QNN circuit design was the density matrix, which contains two totally different sets of complex numbers. I never thought the probability distribution was unchanged for the embedding.'' E4 reported.
\modify{After the second quantum gate RZ gate ( Figure  \ref{fig:case1}\subcomponent{B2})}, E4 further noticed that the line segments representing amplitudes coincide with the semicircles' diameters, while the amplitudes' imaginary parts convert to non-zero values.
% E4 then reported that it is apparent that the embedding for the first gate (\textit{i.e.}, RY gate) does not convert the decimal feature values into the complex number because the imaginary parts are still zero for both quantum states. Next, the RZ gate converts the original number into complex numbers with non-zero real and imaginary parts, as indicated by the line segments in Figure xx.
E4 explained this phenomenon, \textit{``RZ is better to encode information because RZ is for the phase rotation, so this converting makes the input data more resilient to errors due to non-zero imaginary parts.''}
% ``For me, '' E4 then added.
After exploring the two gates for quantum data embedding, 
% E4 was curious about the process of the model layer, which applies the trained model to the embedded quantum states.
% So 
\modify{E4 started to explore the two gates representing the classification model. He first inspected the quantum state after RY gate (Figure \ref{fig:case1}\subcomponent{B3})}.
% selected a data point with two features of 1.4595 and 0.7902, and then 
% output the quantum states after each gate in the model layer.
% As shown in Figure xx, the initial states after the quantum data embedding were converted two times by RY and RZ gates.
\modify{E4 noticed that the probability of state 0 decreased from 0.56 to 0.10, as indicated by the semicircles' area, and almost
% remains 
remained
the same after the last gate (Figure \ref{fig:case1}\subcomponent{B4})}. 
% \yong{1. Pls further check all the tense usage of your writing throughout the paper. 2. When revising some parts, please also read the sentences before and after them to guarantee the logic flow is smooth and makes sense.}
% E4 then found that this is because the absolute values of amplitudes have increased after comparing the corresponding line segments, making the area of the probability of state 0 increase.
\textit{``This is mainly because the absolute values of amplitudes have increased and then cause the area of the probability of State 0 to increase''}, E4 commented.
% \yong{Who said it?}
% E4 then explored the quantum states of several data points with \toolName\ and reported that
% ``This tool is appealing to me because I found no matter how the length of line segments changes, the vertex of the triangle will always locate on the semicircle representing state probability, which makes it much more intuitive to see how the density matrix affects the final measured probability. Integrating the math foundation of quantum computing theory and the visualization is clever and useful.''

\subsection{Case Study \uppercase\expandafter{\romannumeral 2} - Two-qubit Grover's Algorithm}

\begin{figure*}[t]
\centering 
\includegraphics[width=0.87\linewidth]{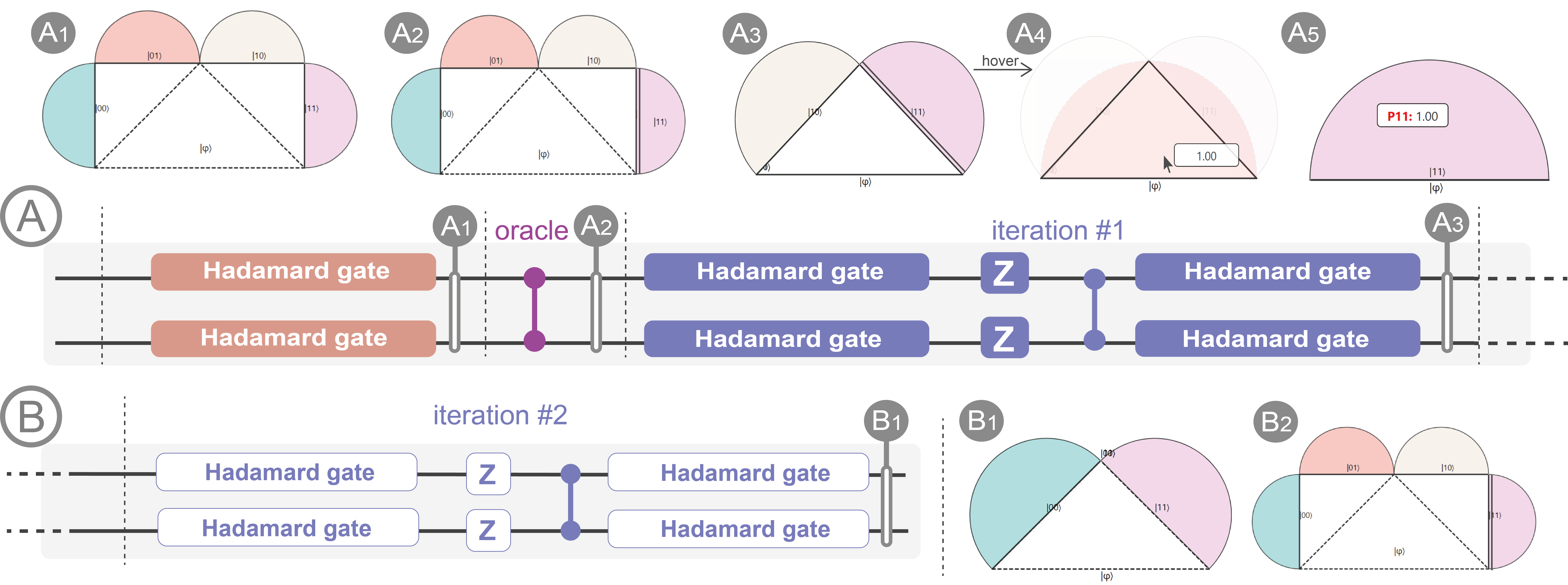}
\caption{
The case for the two-qubit quantum algorithm, \textit{i.e.}, Grover's Algorithm. (A) The calculation process of Grover's Algorithm with one iteration, along with four consecutive quantum states and an interaction shown by \toolName.
(B) The Grover's Algorithm with one more iteration appended after the original circuit, along with two quantum states representation.
}
\label{fig:case2}
\end{figure*}

E9's research interest lies primarily in \textit{Grover's algorithm}~\cite{grover1996fast}, a famous algorithm for the unstructured searching problem.
% , a participant also participated in the expert interview, 
Thus, E9 planned to explore the evolution of quantum states in \textit{Grover's algorithm}.
Following ~\cite{grover}, E9 implemented the circuit with one iteration and the target ``winner'' of state $\ket{11}$.
% The circuit to execute the searching algorithm is shown in Figure \ref{fig:case2}\component{A}.

\textbf{Revealing the insights of the hidden quantum states in Grover's algorithm.}
E9 was curious about the insights of the quantum states provided by \toolName, 
\modify{and expected to see how different modules interact with each other from a view of the functionality block other than individual quantum gates.} 
% \yong{What is ``the systematic circuit''???}
% ``We designed the algorithm from a high-level perspective other than the specific quantum states, so I expect to see something different using this visualization tool.''
Thus, E9 input the density matrix of quantum states into \toolName.
After a glance at the results of the Hadamard gates (Figure \ref{fig:case2}\subcomponent{A1}), 
\modify{which is used to generate the superposition of qubits}, 
% \yong{please check it}
E9 noticed that the probabilities of the four states ($\ket{00}$, $\ket{01}$, $\ket{10}$, $\ket{11}$) are the same, as indicated by the white rectangle at the center and the same area of the four semicircles.
\textit{``It is clear that there is a uniform superposition at this stage because I found the probabilities are all the same. Besides,  the imaginary parts are 0 for all quantum states since the line segments coincide with the semicircle.''}
\modify{E9 then exported the chart after the oracle process (Figure \ref{fig:case2}\subcomponent{A2}) that is used to flip the phase of the searched state.}
% \yong{How can E9 output the chart? He can only download or export the chart.}
% to speculate how the quantum states will be pre-processed before the iteration.
% After comparing the first chart, E9 
He quickly identified a phase flip indicated by the double line for state $\ket{11}$.
Thus, E9 reported, \textit{``Clearly, the quantum state $\ket{11}$ is the one we marked before the initialization due to the negative amplitude shown by the double line.''}
Building upon these findings of the pre-process before the iteration, E9 expected to see \textit{``what the state ``looks like'' after the whole iteration stage.''}
\modify{Hence, E9 exported the visualization of the state (Figure \ref{fig:case2}\subcomponent{A5}) after the diffuser process, which is for the amplitude amplification of the target state. He quickly found that only a purple semicircle remained in the chart.}
\textit{``To my surprise, all semicircles and triangles disappeared compared to the previous figure. This is probably because the iteration found the winner, whose probability is 1.00''}, E9 said.
% \yong{1. Pls carefully check the logic of your claims. 2. Who said this? Please explicitly point it out before this sentence.}
E9 commented, 
\textit{``It is interesting to see the original two-qubit state (\textit{i.e.}, four semicircles) convert to only one semicircle, indicating the searching has converged.''}
% E9 also added that \toolName\ provided him with many insights and details for \textit{Grover's algorithm} analysis based on the intuitive visual evidence instead of the inflexible density matrix.

\textbf{Performing a what-if analysis for multiple iterations.}
\modify{
% Upon the experiment of \textit{Grover's algorithm} with \toolName, 
When using \toolName{} to explore the quantum state evolution of \textit{Grover's algorithm},
E9 was also interested in exploring what would happen if another iteration was implemented on the circuit, as more iteration processes may lead to a better result.
% or remain the same.
}
So he implemented one more iteration block and checked the results using \toolName\ (Figure \ref{fig:case2}\component{B}).
\textit{``As I expected, the searching did not work with another iteration, because the four semicircles with an equal area of 0.25 (Figure \ref{fig:case2}\subcomponent{B2}) unlike the single output in the previous execution} \textit{''}, E9 said.
So, E9 attempted to explore the possible reasons.
\modify{To this end, he compared all charts of each gate in the two iterations and quickly noticed that all states looked the same except those two after the last set of Hadamard gates in the respective iteration (Figure \ref{fig:case2}\subcomponent{A3} and Figure \ref{fig:case2}\subcomponent{B1}).}
Specifically, Figure \ref{fig:case2}\subcomponent{A3} shows that the two states (\textit{i.e.}, $\ket{10}$ and $\ket{11}$) are with the same probability of 0.50 after the first Hadamard gate, while Figure \ref{fig:case2}\subcomponent{B1} indicated that the two states (\textit{i.e.}, $\ket{00}$ and $\ket{11}$) are 0.50 at the same gate during the second iteration.
Then, E9 hovered on the base triangle in Figure \ref{fig:case2}\subcomponent{A3} and the popped tooltip indicated that the probability is 1.00 when the first qubit's state is 1 (Figure \ref{fig:case2}\subcomponent{A4}).
However, no tooltip popped out when he hovered over the second chart.
Thus, E9 concluded that the failure of one more iteration is due to the differences between the two density matrices after the second last Hadamard gate. The last Hadamard gate can convert the state to the one with a probability of 1.00 (Figure \ref{fig:case2}\subcomponent{A4}) with the common first qubit's state with a probability of 0.50. 
% In contrast, the one in the second iteration cannot output the probability of 1.00 because the quantum state does not contain the same first qubit's state (Figure \ref{fig:case2}\subcomponent{B1}). 
\textit{``This is because the Hadamard gate can output the unique value (\textit{i.e.}, 1.00) only if the operated state has the equal value on the first qubit's state.''} E9 reported.
% After the analysis, E9 also gave positive comments that 
% the user interaction impressed him because it provided an intuitive representation when analyzing the single-qubit state in a two-qubit state application. At the same time, it is tough to analyze the density matrix of the quantum state directly.

\section{Expert Interview}
\label{sec:user_study}

% Following the evaluation method proposed by Lee et al.~\cite{lee2022collabally}, w
We conducted in-depth expert interviews with 14 domain experts
to evaluate the performance of \toolName\ for given tasks.
% Overall, our method used for the user study is task-based usability testing.
% We expect to understand the strengths and limitations of \toolName\ for quantum state representation tasks.

\vspace{-0.3cm}
\subsection{Participants and Apparatus}

We invited 14 experts (2 females, 12 males) in quantum computing to join our well-designed interviews.
Participants ($age_{mean}= 33.07$, $age_{sd}= 6.16$) were with an average of over five years of quantum computing research experience.
Specifically, E4-8 are working on Quantum Machine Learning;
E9-12 current research direction is Quantum Error Modeling; 
E13-14 are studying Quantum Chemistry, while the other three participants study Quantum Compiler (E1), Quantum Systems (E2), and Quantum Simulator (E3), respectively. 
Meanwhile, E3 and E12 are from the same research institute in the U.S., while others work in different educational institutions in the U.S.
The above participants differ from the domain experts involved in the co-design process.
% To guarantee the findings collected in the expert interviews are general for common users, 
% and none of them has a background in Visualization or HCI.
All participants were asked to use a monitor with a resolution of 1920 $\times$ 1080 in advance.

% Our study was reviewed and approved by our institution's IRB.

% In our study, E1 and E2 used our baseline visualization approach, \textit{i.e.}, Bloch Sphere, while E3-14 used \toolName\ as the apparatus.
% Apart from the different apparatus, all participants were given the same tasks described in Section \ref{sec:tasks}.
% We first report the results of E1 and E2 using Bloch Sphere in Section \ref{sec:result} to report the current challenges and limitations in quantum state observation using Bloch Sphere.
% We then present the results from the remaining 12 participants (E3-14), who used our visualization approach \toolName.

\subsection{Procedure}

Before the formal study began, participants were asked questions about their demographics and research directions.
% We then .
We first introduced the goal and detailed procedures of the study, and then showcased an example to illustrate the interface and usage of \toolName.
% including both single-qubit and two-qubit state representation.
The above process lasted about 20 minutes.

% Participants then accessed the \toolName\ package on their laptops or desktops.
After the introduction, they were expected to accomplish the pre-defined tasks described below in Section \ref{sec:tasks}, which were about to observe the qubit state using the \toolName\ interface.
Note that the application to perform the qubit state visualization was in line with the research direction of each participant so that we can evaluate the generalizability of \toolName\ for different domain tasks in quantum computing (\textit{e.g.}, quantum algorithm, quantum machine learning).
% The research directions of each participant are shown in Table \ref{tab:1}.
We recorded and took notes on each task and participants' interaction processes.
The above process lasted about 45 minutes.

Upon task completion, all participants were then invited to participate in the post-study interview.
Each participant was encouraged to describe the advantages and limitations in a think-aloud manner.
% Specifically, they provided feedback regarding how \toolName\ can visually support the analysis of quantum state vector, the probability distribution of quantum states, and quantum entanglement visualization.
Meanwhile, they were also encouraged to describe the issues of Bloch Sphere regarding these three aspects and how these issues affect their routine tasks.
We invited every participant to rate the \toolName\ using a 7-Likert scale based on the questionnaire (Table \ref{tab:1}).
The above interview and rating procedures took 30 minutes.
\subsection{Tasks}
\label{sec:tasks}

We designed the tasks to mimic the everyday tasks for analyzing the quantum states.
To enhance the generalizability and make it not limited to one single application in quantum computing, we asked each participant to complete the tasks based on a typical example in their research directions (\textit{e.g.}, variational quantum circuit for Quantum Machine Learning, QEC~\cite{lidar2013quantum} algorithm for Quantum Error Modeling, or VQE~\cite{kandala2017hardware} algorithm for Quantum Chemistry).  
We proposed three types of tasks as follows:

\begin{table}[t]
  \caption{
The questionnaire provided during the expert interview based on the tasks in the expert interview, \textit{i.e.},
Task 1 (\textbf{Q1-2}), Task 2 (\textbf{Q3-4}), Task 3 (\textbf{Q5-6}), and an overall feedback of \toolName\ (\textbf{Q7-9}).
  }
  \label{tab:1}
\begin{tabular}{p{0.4cm}|p{7.5cm}}
\hline
Q1  & It is useful to show the quantum entanglement when observing quantum states.                         \\
Q2  & It is easy to identify the entangled states via the visually correlated semicircles.                 \\ \hline
Q3  & It is helpful to show the probability distribution.                              \\
Q4  & It is intuitive to show probability distribution via the semicircle area.                            \\ \hline
Q5  & It is informative to represent states via the state vectors.                                 \\
Q6  & It is easy to identify the state vectors via the line pairs within each semicircle.                  \\ \hline
% Q7  & Overall, \toolName\ can better visualize qubit states than Bloch Sphere. \\
Q7  & The user interactions in the interface are useful and smooth.                                        \\
Q8  & The design can be integrated into the workflow well.                                                 \\
Q9 & The 2D visual design is easy to view.                                                                \\ \hline
\end{tabular}
\end{table}

\begin{figure}[t]
\centering 
\includegraphics[width=\linewidth]{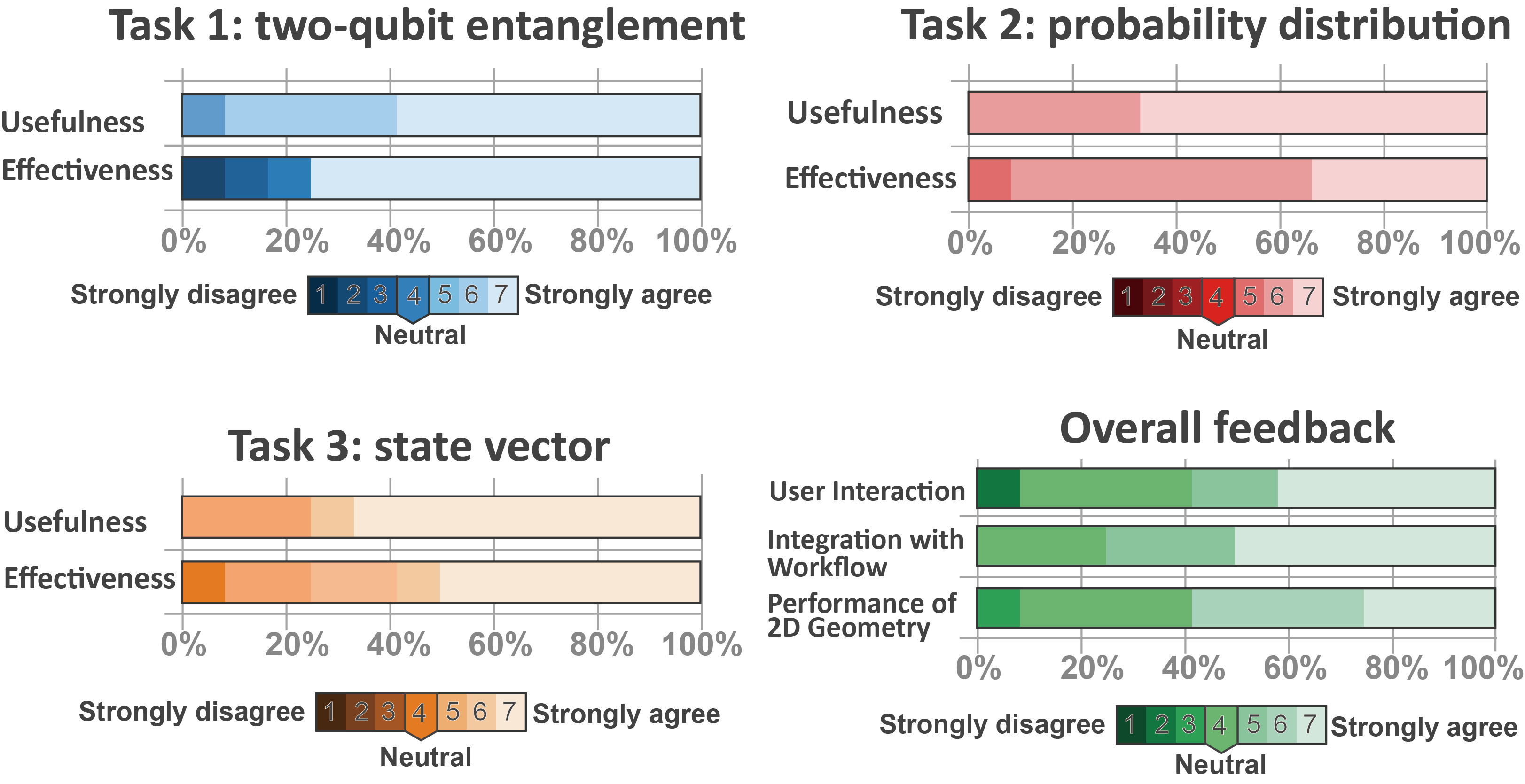}
\caption{The summary of the user feedback, which consists of the results of the three analytical tasks and the overall feedback.}
\label{fig:4}
\end{figure}

\textbf{Task 1: Explore the two-qubit entanglement.}
Participants were asked to identify all two-qubit states in \toolName, describe the components in each two-qubit state, analyze the single-qubit state in the two-qubit quantum system states
% (using the interaction of ``display order'' in the control panel)
, and describe how this feature benefits their quantum state analysis.

\textbf{Task 2: Observe the state probability distribution.}
Participants were asked to identify the probability of each quantum state and compare all quantum state probabilities via the corresponding semicircle area. User interactions displaying the exact probability amplitudes are allowed after the answer is given.

\textbf{Task 3: Identify the state vector.}
Participants were asked to identify the real and imaginary parts of each state vector via the line segments on the right triangles and describe how the length of the line segments affects the semicircle area.
% User interactions are only allowed if the line segment coincides with the diameter of the semicircle.

\section{Interview Results}
\label{sec:result}

In this section, we report the summarized results, including the Bloch Sphere's challenges, the responses for the three analytical tasks, and the suggestions provided by the participants. We reported the ratings from each participant in Figure \ref{fig:4}.

\subsection{Current Bloch Sphere Challenges}
We invited all participants to participate in the feedback collection for Bloch Sphere, collecting challenges faced by the participants when attempting to perform quantum state observation using Bloch Sphere. We use this session to confirm our previous findings from the co-design process.

\textbf{Challenge 1: Incapability of the two-qubit state representation.}
All participants agreed or strongly agreed that the inability to support two-qubit representation is a severe limitation for Bloch Sphere, they all agreed that it is unacceptable that Bloch Sphere cannot support multiple qubits' observation, which is the most critical property for quantum computing.
E7 also reported that Bloch Sphere needs to be more scalable to allow the visualization of quantum entanglement.

\textbf{Challenge 2: Non-intuitiveness of the state probability distribution.}
Most participants (11 out of 14) confirmed that Bloch Sphere could not intuitively visualize the probability of each possible quantum state.
E13 mentioned that he has to manually calculate the final measurement from the point's location in Bloch Sphere. However, it is not a trivial task \textit{``because the rotation angle in each axis of Bloch Sphere is hard to measure in a 3D model.''}

\textbf{Challenge 3: Lack of the state vector representation.}
Most participants (10 out of 14) agreed that the missing state vector in Bloch Sphere makes it challenging to understand.
E1 commented, \textit{``Despite the transparency of rotation gates for Bloch Sphere, but
I prefer a more intuitive way to reflect the density matrix
when designing  circuits with  quantum simulators.''}
E8 added that Bloch Sphere is unsuitable for educational purposes because of its complexity, especially for entry-level users.

\subsection{Results of Three Analytical Tasks}

We summarized all feedback regarding the three analytical tasks shown in Section \ref{sec:tasks}.

\textbf{Task 1: Two-qubit entanglement analysis.}
All participants agreed or strongly agreed that it was easy to support the two-qubit state observation using \toolName\ \textbf{($rating_{mean}= 6.01$, $rating_{sd}= 1.51$)}.
% E14 emphasized \textit{``I like the idea of having more than one qubit and being able to represent their states, on the Bloch sphere that's not so helpful, adding more qubits seems important to broaden its potential.''}
% Also, all participants confirmed that the feature of two-qubit state representation could yield significant benefits for the visual entanglement analysis.
Specifically, E3 believed that \toolName's most practical usage is to \textit{``make it possible for the entanglement display, which is the most important characteristic and needs for every quantum computing researcher.''}
% Most participants (9 out of 14) praised that \toolName\ enables the two-qubit state visualization using the same geometries for single-qubit visualization, which \textit{``makes the visualization more accessible for general quantum computing users.'' }
% For the specific research directions in quantum computing,
% all participants agreed that \toolName\ could practically assist them in their research domains in quantum computing.
% For example, 
E14, an expert in Quantum Chemistry, mentioned, \textit{``my research is mainly focused on cutting the large quantum circuit; usually, the circuit has more than two qubits. This visualization of more qubits will help me to
% better understand the deeper structure of a large circuit and will help me to 
find the optimal cutting point more accurately.''}
Furthermore, as an expert on Quantum Error Modeling, E9 noted, \textit{``I can see a great potential of \toolName. I think the quantum error correction (QEC) research can make good use of \toolName, for visualizing the encoding qubits and ancilla qubits separately.''}

\textbf{Task 2: State probability observation.}
Most participants agreed that they could identify the probability of all possible quantum states quickly ($rating_{mean}= 6.60$, $rating_{sd}= 1.42$).
E12 mentioned that the encoding area to visualize probability is
\textit{``easy and straightforward''}.
E4 praised that, \textit{``I like the idea of using the probability calculation equation to naturally visualize \modify{the} probability distributin. I can directly check the probability without any manual calculation.''}
% E4, who was working on Quantum Machine Learning and used Bloch Sphere frequently, commented, \textit{``I believe that it gives the user a more intuitive way to observe the probability compared to Bloch Sphere.''}
For the specific research directions in quantum computing,
most participants (9 out of 12) highly appreciated \toolName's usability and felt it could help them to handle their domain-specific tasks smoothly.
% E14 commented, \textit{``this tool helps me understand quantum circuit cutting method, especially this probability distribution provides me a better view for finding an optimal cutting point on a quantum circuit.''}
For example, E4 confirmed that \textit{``(for Quantum Machine Learning) \toolName\ can easily visualize the probability at breakpoints in  debugging, which saves time in in-line debugging.''}

\textbf{Task 3: State vector exploration.}
Most participants agreed that it was intuitive to get a sense of the state vectors in detail using \toolName\ ($rating_{mean}= 5.85$, $rating_{sd}= 1.24$).
Specifically, 8 out of 14 participants felt it helpful to show the amplitudes (\textit{i.e.}, $\alpha$ and $\beta$) separately using multiple right triangles.
% E12 said it was practical to represent quantum states using the state vectors because it makes domain users more easily understand the constructions of quantum states.
E9 agreed that it provides a better view of the relationships between the probability distribution, especially the entanglements.
Due to the accessibility of state vectors, E11 thought the visualizations of the real and imaginary parts of amplitudes were helpful \textit{``for fresh starters in quantum computing.''}
% Participants also commented that the state vector representation could be well integrated into their daily tasks for the specific research directions in quantum computing.
% E14 reported that knowing $a$ and/or $b$ can help him better visualize and understand the measurement basis for quantum circuit cut: \textit{``depending on the value of a and/or b, I can choose a proper measure basis for the post-processing step.''}
Moreover, E2 confirmed the \toolName's value in Quantum System, \textit{``I can quickly compare the consequence of using different gates as it can fit into the debugger.''}
% He also felt that \toolName\ could significantly reduce the requirements of developers to understand quantum physics, making it an excellent in-class teaching tool for non-physics background students.

\vspace{-0.3cm}

\subsection{Overall Feedback}
From the participants' responses and ratings (\textit{i.e.}, $rating_{mean}= 5.93$, $rating_{sd}= 1.03$), we summarize the overall feedback regarding the user interaction, the integration with the workflow, and the performance of 2D geometry.
\textit{1) User interactions.}
Participants agreed that the user interactions of \toolName\ are valuable and smooth. They enjoyed the overall interactions during the study. Among them, participants gave highly positive feedback for the feature of the switching of qubit display orders.
\textit{2) Integration with the workflow.}
Participants liked how \toolName\ can seamlessly fit into their specific domain tasks.
Building upon the publicly-available platform, \toolName\ can benefit all quantum computing developers and researchers.
% Besides, we open-sourced the code to enable any learning and extension for the current visualization.
\textit{3) Performance of 2D geometry.}
Participants agreed that using 2D shapes was more readable and apparent than 3D approaches.
For instance,
% E6 pointed out that \toolName\ flattens the states and makes the perception of each characteristic more accurate than 3D views.

\section{Discussion}

In this section, we first summarize the lessons we learned during the development of \toolName, 
then we discuss the limitations of \toolName.

\vspace{-0.3cm}

\subsection{Lessons}
% We learned many valuable lessons from the design and evaluation of our visual design.

% \textbf{Fitting visualization into quantum computing.}
During the above evaluation processes, all participants gave highly positive feedback for \toolName.
Among all the responses, participants emphasized a strong need for visualization to fit into quantum computing regarding the complex quantum physics theory, non-transparency of quantum program process, and non-intuitive quantum computing properties (\textit{i.e.}, quantum entanglement and superposition).
The above challenges make it hard for novices and the general public to have a strong sense of quantum computing.
Thus, the quantum computing community urgently needs visualization to aid the transparency and interpretability of quantum computing with its scientific educational capability.

% \textbf{Lowering the learning curve of visual designs for quantum computing users.}
% During the co-design process, five domain experts strongly emphasized the difficulty for quantum computing users to learn and use cross-disciplinary visualizations. 
% All domain experts preferred concise and straightforward visual designs, which can really help them instead of those sophisticated visual solutions.
% Thus, \toolName\ received highly positive feedback concerning the simplicity and intuitiveness, giving it the potential to be widely spread by general users in quantum computing.
% Furthermore, we learned from the interviews that it is crucial to correlate the visual elements using quantum computing characteristics other than using many different and individual visual channels, such as the visual solution provided by \textit{Quirk}.

\vspace{-0.4cm}

\subsection{Limitations}

Our evaluation shows that \toolName\ can effectively facilitate quantum state observation. However, there are still some limitations.

\textbf{Limited support for quantum noise visualization.}
\toolName\ can effectively visualize various quantum states in situations where noise analysis is not required,
such as the design and debugging of quantum algorithms. 
% \yong{why? No noise in these applications??}
\modify{We do not consider the noise analysis of  \toolName\ because the design is built upon quantum simulators where the execution of quantum circuits is completely noise-free.}
% \modify{We do not consider the noise model of the simulators since too complex functionality might confuse the general users and learners under the education scenarios.}
% It would be better to support the comparison with noisy and noise-free quantum state results.
% \yong{Pls check my comments.}

\textbf{Scalability.} 
% As the current focus of \toolName\ is for 1-qubit and 2-qubit problems,
\toolName{} currently targets visualizing the quantum states of one or two qubits. Compared with Block Sphere, it can effectively visualize the quantum entanglement, a significant step towards effective qubit state visualization confirmed by the participants.
% Meanwhile, there are requests to prepare for the next generation of quantum computers with a scale difference of more than an order of magnitude. 
Also, E8 suggested enabling representation for more qubits by adding more triangles on top of \toolName. In the future, we will endeavor to extend the current design for more qubits.

\textbf{Time-consuming input for state vectors.}
The update of \toolName\ is driven by the inputted number of amplitudes, which requires users to input the real and imaginary parts of amplitudes manually.
Participants reported that it is inconvenient to type in the amplitude values.
However, due to our contribution to a design study, we plan to address this limitation in the future.
For example, as hinted by E4, converting from the popular visualization, Bloch Sphere, will also be helpful for users.

\section{Conclusion}

We present \toolName, a novel visualization approach for quantum state representation.
Built upon the visualization of the state vectors, we utilized multiple coordinated 2D geometrical shapes to reflect the probability distribution of all quantum states, which provides an intuitive view of quantum superposition.
Also, we applied the same form of visualization approaches to enable \toolName\ to support single-qubit and two-qubit state representations.
% , where two-qubit states include two-qubit entanglement.
We evaluated \toolName\ by conducting two case studies and in-depth expert interviews, and
the results show that \toolName\ can effectively facilitate the visual analysis of single-qubit and two-qubit states.

In the future, we plan to enhance the data loading of quit states by making the user interface of \toolName\ able to read qubit states from quantum computing circuits automatically.  
Also, it will be interesting to investigate how visualization techniques can be applied to show the quantum states of more than two qubits. 

% focus on enhancing the functionality of qubit state representation, like integrating the quantum noise representation into \toolName\ via multiple glyphs, which can highlight the differences between the noisy and noise-free results for post-execution analysis.
% Furthermore, we could enhance the efficiency of \toolName\ exploration, such as a web browser plugin that can recognize and visualize the quantum states automatically.

\section*{Acknowledgment}

This work was supported by the Lee Kong Chian Fellowship awarded to Dr. Yong Wang by SMU. Qiang Guan was supported by NSF 2212465, 2230111, 2217021 and 2238734.

\bibliographystyle{eg-alpha-doi}  
\bibliography{egbibsample}

\end{document}